\documentclass[pr, twocolumn, preprintnumbers, showpacs, superscriptaddress,longbibliography, footnoteadded]{revtex4-1}
\pdfoutput=1

\usepackage[T1]{fontenc}
\usepackage{amssymb}
\usepackage{graphicx}
\usepackage{amsmath}
\usepackage[
      colorlinks=true,
      linkcolor=blue,
      urlcolor=blue,
      filecolor=black,
      citecolor=blue,
            ]{hyperref}
\usepackage{tensor}
\usepackage{epstopdf}
\usepackage{color}

\newcommand{\td}{\text{d}}
\def\mO {\mathcal{O}}
\def\mo {\langle\mathfrak{o}\rangle}
\def\moi {\langle\mathfrak{o}_i\rangle}

\def\g {\mathfrak{g}}
\def\mphs {\varphi^{(s)}}
\def\mphe {\varphi^{(e)}}

\begin{document}
\title{On the applicability of holography in thermodynamic equilibrium}

\author{Run-Qiu Yang }
\email{aqiu@tju.edu.cn}
\affiliation{Center for Joint Quantum Studies and Department of Physics, School of Science, Tianjin University, Yaguan Road 135, Jinnan District, 300350 Tianjin, China}

\author{Li Li}
\email{liliphy@itp.ac.cn}
\affiliation{CAS Key Laboratory of Theoretical Physics, Institute of Theoretical Physics,
Chinese Academy of Sciences, Beijing 100190, China}
\affiliation{School of Fundamental Physics and Mathematical Sciences, Hangzhou Institute for Advanced Study, UCAS, Hangzhou 310024, China}
\affiliation{Peng Huanwu Collaborative Center for Research and Education, Beihang University, Beijing 100191, China.}

\author{Rong-Gen Cai}
\email{cairg@itp.ac.cn}
\affiliation{CAS Key Laboratory of Theoretical Physics, Institute of Theoretical Physics,
Chinese Academy of Sciences, Beijing 100190, China}
\affiliation{School of Fundamental Physics and Mathematical Sciences, Hangzhou Institute for Advanced Study, UCAS, Hangzhou 310024, China}

\begin{abstract}
For a strongly coupled system that has a gravity dual description, we show that the standard holographic dictionary yields a nonnegative susceptibility when the system is in thermodynamic equilibrium and the correlation function is absolutely integrable.  When the system has no spontaneous condensation or has a spontaneous $\mathbb{Z}_2$-symmetry breaking, we find that the ``trace energy condition'' is violated in many cases (see Eq.~\eqref{tracebk}). There is a normalized grand potential density that is monotonic as accessing to lower scales, providing a candidate $c$-function characterizing the number of effective degrees of freedom. Finally, we discuss a ``paradox'' raising by the negative susceptibility in holography and its resolution.
\end{abstract}
\maketitle
\flushbottom

\noindent

\section{Introduction}\label{intro}
Strongly coupled systems are ubiquitous in Nature, ranging from nuclear physics, fluid dynamics, astrophysics to condensed matter, etc.  As the traditional perturbation approach ceases to be applicable, it has been challenging to understand those systems that involve strong interaction in the non-perturbative regime. In recent years, the holographic duality~\cite{Maldacena:1997re,Gubser:1998bc,Witten:1998qj,Witten:1998zw}, which origins from string theory, offers us a prospective tool to crack this hard nut. By mapping a $d$-dimensional strong coupling theory to a $(d+1)$-dimensional asymptotically anti-de Sitter (AdS) spacetime, one can instead deal with generic gravitational phenomena in terms of classical general relativity.  This holographic approach has been used to study various strongly coupled systems, such as quark-gluon plasma~\cite{Policastro:2001yc,Erdmenger:2007cm,Aoki:2012th,Kovtun:2004de,Herzog:2006gh}, high temperature superconductivity~\cite{Hartnoll:2008vx,Gubser:2008wv,Cai:2013aca,Kim:2013oba,Zaanen:2021llz},  strange metals~\cite{Hartnoll:2009ns,Sachdev:2010uj,Myers:2010pk,Sachdev:2012dq,Baggioli:2022pyb} and fermi/non-fermi liquids~\cite{Liu:2009dm,Faulkner:2009wj} and so on.

Since intensive investigations have been made, a natural question arises: what kind of strongly coupled systems has a dual gravitational description in holography? Considering a dual system to be a critical point where an exact conformal symmetry emerges, the authors of~\cite{Kovtun:2008kw} provided a criterion on whether such a critical point admits a dual gravitational description. They argued that the normalized entropy density (defined in Eq.~\eqref{defcs1}) should be equal to the central charge. However, in most applications of holographic duality or interesting strongly coupled systems, the boundary theory is deformed by some relevant operators. In these cases, the criterion of~\cite{Kovtun:2008kw} cooked for a conformal field theory (CFT) fails. As we will show explicitly, the normalized entropy can be different from the central charge at the UV fixed point.

In this work, we give some judgments on whether a strong coupling system has a dual description in holography based on general considerations.  For a $d$-dimensional  quantum field theory at finite temperature  deformed by $\mathcal{N}$ relevant operators $\mathfrak{o}_i$ with scaling dimension $\Delta_i$ ($i=1,2,\cdots,\mathcal{N}$), its thermodynamics is governed by the grand potential $\psi=\psi(T,\vec{J})$ such that $\moi=-\partial\psi/\partial J_i$. Here $J_i$ and $\moi$ are thermodynamic conjugate variables, for which $\moi$ are typically the relevant conserved charges of the system (\emph{e.g.} the electric charge density) and $J_i$ the corresponding ``chemical potentials''. By using the basic holographic dictionary, we will prove that the generalized susceptibility $\partial\moi/\partial J_i$ for a system with a gravity dual should be nonnegative when the correlation function is absolutely integrable.  Moreover, we will show the violation of the trace energy condition for a thermodynamic stable state, and introduce a normalized grand potential density that is monotonic as accessing to lower scales, providing a candidate c-function characterizing the number of effective degrees of freedom at given energy scales. Our results give a strong constraint on whether a gravitational theory can describe any lower-dimensional thermal equilibrium system within the holographic duality. Finally, we will discuss a ``paradox'' rasing by the negative susceptibility appearing in holography and its resolution.

\section{Review on holography}
Referring to the holographic duality, the operators $\mathfrak{o}_i$ of boundary field theory are dual to bulk fields $\{\varphi_i, i=1,2,\cdots\mathcal{N}\}$ in one higher dimension. The gravitational bulk theory is given by the following action.
\begin{equation}\label{bulkaction}
  {S}=\int\td^{d+1}x\sqrt{-g}\left[R+\frac{d(d-1)}{\ell_{\text{AdS}}^2}+\mathcal{L}_m\right]+{S}_{ct}\,.
\end{equation}
Here $g$ is the determinate of the bulk metric, $R$ is the scalar curvature, $\mathcal{L}_m$ stands for the Lagrangian of matter sector, and ${S}_{ct}$ denotes some boundary terms which cancel the UV divergence and ensure the variation well-defined. We will set the AdS radius $\ell_{\text{AdS}}=1$ with $16\pi G_N=c=\hbar=k_B=1$ in our following discussion.
By choosing a suitable coordinate system with the holographic radial coordinate $r$,  the asymptotical expansion of each matter field near the AdS boundary at $r\rightarrow\infty$  has two independent branches.
\begin{equation}\label{aspseries1}
  \varphi_i=\mphs_ir^{\tilde{\Delta}_i-d+s_i}(1+\cdots)+\varphi_i^{(e)} r^{s_i-\tilde{\Delta}_i}(1+\cdots)\,,
\end{equation}
where $s_i$ is the rank of $\varphi_i$~\footnote{We here only consider bosonic fields and assume that fields of $s_i\geq1$ are $s_i$-form fields which in String/M-theory are well-defined as the sources of the brane.}. Without loss of generality, we assume $d-\tilde{\Delta}_i\leq\tilde{\Delta}_i$ such that $\tilde{\Delta}_i\geq d/2$. In the so called ``standard quantization'', one considers the leading term $\vec{\varphi}^{(s)}=(\mphs_1,\cdots,\mphs_{\mathcal{N}})$ as the sources $\vec{J}$ of the dual system. In this case, the scaling dimension $\Delta_i$ of $\mathfrak{o}_i$ is $\Delta_i=\tilde{\Delta}_i$ and the scaling dimension of the source $\mphs_i$ is $d-\tilde{\Delta}_i$. One may have the so-called alternative quantization by choosing $\mphe_i$ as the source, for example, for the scalar case with $d/2\leqslant\tilde{\Delta}_i\leqslant(d+2)/2$.

According to the standard holographic dictionary~\cite{Gubser:1998bc,Witten:1998qj}, turning on the external source $\mphs$ of the bulk field $\varphi$ corresponds to introducing the deformation $\int\mphs_i\mathfrak{o}_i\td^dx$ for an operator $\mathfrak{o}_i$ in the dual field theory. In the thermodynamical equilibrium case, the precise relationship is given by the identification of the Euclidean partition functions for both the bulk and field theories (We follow the convention of Ref.~\cite{Witten:1998qj})~\footnote{There is also a different convention in the literature, see \emph{e.g.}~\cite{Skenderis:2002wp}}:
\begin{equation}\label{gkpw}
 Z_{\text{QFT}}= \left\langle\exp\sum_i\int\mphs_i\mathfrak{o}_i\td^dx\right\rangle_{\text{QFT}}=Z_{\text{bulk}}[g_{\mu\nu}^{(E)},\varphi_i]\,,
\end{equation}
where the bulk partition function is computed with the boundary condition that at the asymptotically AdS boundary $\varphi_i$ approaches to a given source term $\mphs_i$ of Eq.~\eqref{aspseries1}. In the leading saddle point approximation, one can compute $Z_{\text{bulk}}[g_{\mu\nu}^{(E)},\varphi_i]$ by  the on-shell bulk action ${S}_{\text{Euclidean,on-shell}}$, \emph{i.e.} $Z_{\text{bulk}}=e^{-{S}_{\text{Euclidean,on-shell}}}$. On the other hand, for a system in thermodynamic equilibrium that is described by a stationary black hole with a well-defined temperature $T$, the standard black hole thermodynamics yields that the free energy (grand potential) $\Psi$ is given by $\Psi=-T\ln Z_{\text{bulk}}= T{S}_{\text{Euclidean,on-shell}}$. We will focus on a  homogeneous system in flat spacetime. Denoting $\Omega_{d-1}$ to be the spatial volume of the dual theory, the free energy density reads
\begin{equation}
 \psi=\frac{T}{\Omega_{d-1}}{S}_{\text{Euclidean,on-shell}}=-\frac{T}{\Omega_{d-1}}\ln Z_{\text{QFT}}\,.
 \end{equation}

By definition, the expectation value of any operator $X$ in the above thermodynamic system is given by
\begin{equation}\label{expact10}
  \langle X\rangle:=Z_{\text{QFT}}^{-1}\left\langle X\exp\sum_i\int\mphs_i\mathfrak{o}_i\td^dx\right\rangle_{\text{QFT}}\,.
\end{equation}
Since the external source $\mphs_i$ contributes to the partition function via Eq.~\eqref{gkpw}, one can prove
\begin{equation}\label{expact1}
  \moi=-\partial\psi/\partial\mphs_i\,.
\end{equation}
See appendix~\ref{expectval} for more details. When $\mphs_i\neq0$, one should require $\tilde{\Delta}_i<d$ such that the operator $\mathfrak{o}_i$ is relevant (or equivalently, the source will not destroy the asymptotically AdS geometry as $r\rightarrow\infty$).\\

\section{Nonnegativity of susceptibility}
The basic holographic dictionary requires that the external source $\mphs_i$ contributes to the partition function according to Eq.~\eqref{gkpw}. The first-order derivative of $ Z_{\text{QFT}}$ with respect to the source gives Eq.~\eqref{expact1}. What will one obtain if considering the second-order derivative? To answer this question, let us rewrite the partition function of the homogenous thermal equilibrium system into the following form.
\begin{equation}\label{homoz1}
  Z_{\text{QFT}}=\langle e^{T^{-1}\Omega_{d-1}\sum_i\mathfrak{o}_i\mphs_i}\rangle_{\text{QFT}}=\langle e^{T^{-1}\sum_i\mO_i\mphs_i}\rangle_{\text{QFT}}\,,
\end{equation}
with $\mO_i:=\mathfrak{o}_i\Omega_{d-1}$. Then we have~\footnote{It needs to note that the expectation value of $\mathfrak{o}_i$ depends on $\mphs_i$ but operator $\mathfrak{o}_i$ itself in Eq.~\eqref{gkpw} is a variable independent of $\mphs_i$.}
\begin{equation}\label{homoz2}
  \moi=T\Omega_{d-1}^{-1}\partial_{\mphs_i}\ln Z_{\text{QFT}}\,,
\end{equation}
and
\begin{equation}\label{homoz3}
  T\Omega_{d-1}\frac{\partial\moi}{\partial\mphs_i}=T^2\partial^2_{\mphs_i}\ln Z_{\text{QFT}}=\langle\mO_i^2\rangle-\langle\mO_i\rangle^2\,.
\end{equation}
This leads to
\begin{equation}\label{sucepz1}
  \frac{\partial\langle\mathfrak{o}_i\rangle}{\partial\mphs_i}=\frac{\langle(\mO_i-\langle\mO_i\rangle)^2\rangle}{T\Omega_{d-1}}\geq0\,.
\end{equation}
A similar result can be obtained in inhomogeneous states, see appendix~\ref{suscep} for more details. One sees that the special coupling required in Eq.~\eqref{gkpw} not only leads to Eq.~\eqref{expact1} but also implies a ``fluctuation-susceptibility relation''~\eqref{sucepz1}. This is a direct corollary of the basic holographic dictionary, Eq.~\eqref{gkpw}, but did not draw sufficient attention in the literature. At the end of this paper, we will show that some widely used bulk models do not match Eq.~\eqref{sucepz1}. In addition,  let us stress here that the susceptibility could be negative if the external source $\mphs_i$ contributes to the partition function in a different way (for example, the diamagnetic materials, see appendix~\ref{magsuscp} and Ref.~\cite{9780123821881} for more discussions).

We emphasize that the nonnegativity of susceptibility results from thermodynamic stability is only valid in a few special cases, for example, the ``heat capacity'' (the susceptibility of temperature) and the  ``minus of compressibility'' (the susceptibility of pressure). For more general cases they have no relationship with each other and thus one cannot use thermodynamic stability to argue the nonnegativity of susceptibility.  One can refer to appendix~\ref{suscepstab} for more detailed discussions.

\begin{figure}
\centering
\includegraphics[width=0.20\textwidth]{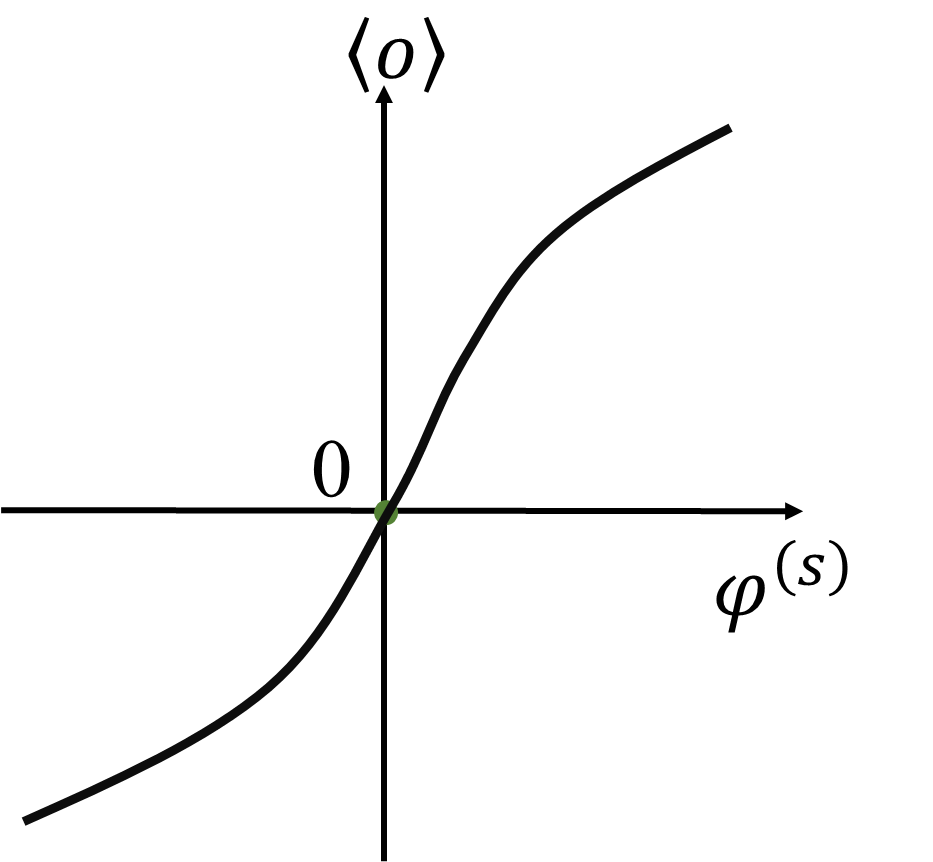}
 \includegraphics[width=0.20\textwidth]{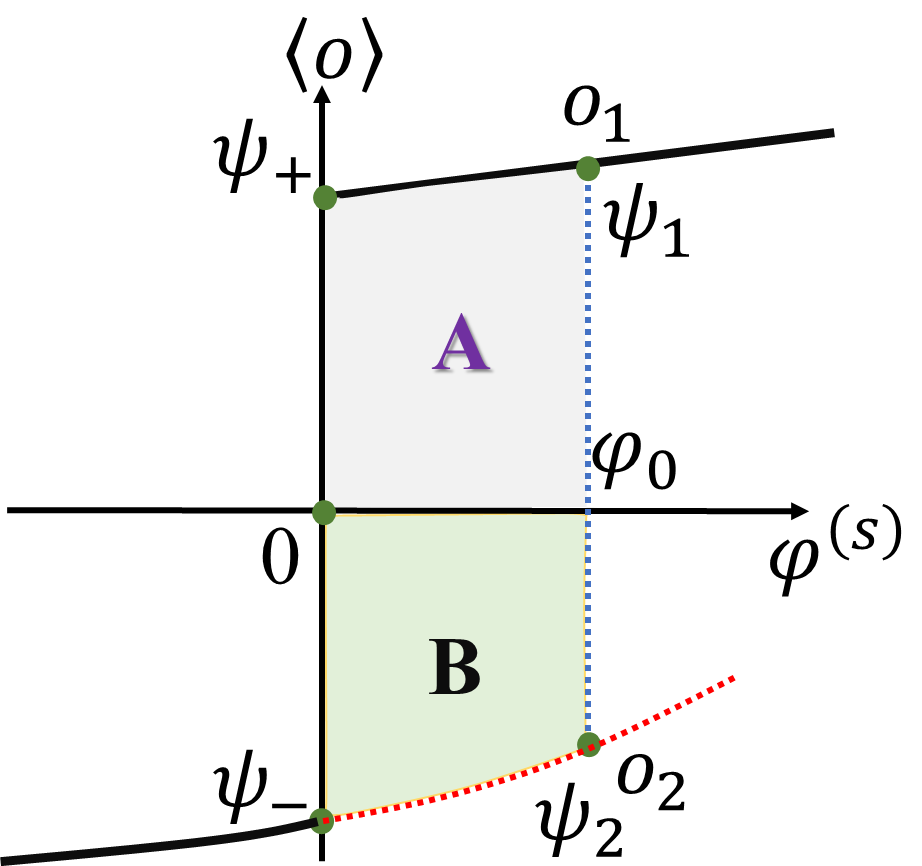}
   \caption{The condensation $\mo$ with respect to the source $\mphs$. Left panel: no spontaneous condensation. Right panel: spontaneous $\mathbb{Z}_2$-symmetry breaking at $\mphs=0$. The black solid lines stand for the thermodynamically favored trajectory.}\label{figophi}
\end{figure}

We now show that Eq.~\eqref{sucepz1} gives the following constraint
\begin{equation}\label{os}
\moi\mphs_i\geq0\,.
\end{equation}
for a thermodynamically stable phase if it has no spontaneous condensation or has spontaneous $\mathbb{Z}_2$-symmetry breaking when $\mphs_i=0$, shown schematically in Fig.~\ref{figophi}. Although the latter does not describe all cases with spontaneous symmetry breaking, a large class of interesting phenomena including superconductivity and (anti)ferrimagnetism, have the spontaneous $\mathbb{Z}_2$-symmetry breaking.

In the first case, the system has no spontaneous condensation when $\mphs_i=0$, \emph{i.e.} $\moi=0$ if $\mphs_i=0$ (left panel of Fig.~\ref{figophi}). Then Eq.~\eqref{sucepz1} immediately implies $\mphs_i\moi\geq0$.
In the second case, the system has spontaneous condensation, \emph{e.g.} there is a critical temperature $T_c$, below which the \textit{thermodynamically favored states} have $\moi\neq0$ when $\mphs_i=0$. As can be seen from the right panel of Fig.~\ref{figophi}, at $\varphi^{(s)}=0$ there are two condensed phases with the free energies $\psi_-=\psi_+$ lower than that of the uncondensed phase. When $\varphi^{(s)}=\varphi_0>0$ (the case for $\varphi_0<0$ is similar), since the susceptibility~\eqref{sucepz1} is nonnegative, there can be two branches  of $\mathfrak{o}$ labeled by $o_1$ and $o_2$, respectively. Form Eq.~\eqref{expact1}, one can find that their grand potential densities satisfy $\psi_1-\psi_+=-$area of ``gray region A'' and $\psi_2-\psi_-=$area of ``green region B''. Therefore, the state corresponding to $o_1$ has lower free energy and thus is thermodynamically favored. Moreover, this thermodynamically favored state has $\langle\mathfrak{o}\rangle\varphi_0=o_1\varphi_0>0$. Therefore, we can conclude that $\mphs_i\moi\geq0$ in a physically favored state.  Note that here we do not consider the metastable states. \\

\section{Violation of trace energy condition}
Using the standard holographic dictionary~\cite{Skenderis:2002wp,Balasubramanian:1999re,Bianchi:2001kw}, the energy momentum tensor $\tau_{ab}$ of the dual field theory is given by
\begin{equation}\label{tabm}
  \tau_{ab}=-\lim_{r\rightarrow\infty}\frac2{\sqrt{-h}}\frac{ \delta{S}|_{\text{on-shell}}}{\delta h^{ab}}\,.
\end{equation}
Here $r^2h_{ab}$ is the induced metric of the AdS boundary and $h_{ab}|_{r\rightarrow\infty}=\eta_{ab}$ is the metric of the dual boundary theory.  We now show that, if the systems have no Weyl anomaly, the energy-momentum tensor of Eq.~\eqref{tabm} will satisfy
\begin{equation}\label{bdtabb2}
  \tau=\sum_{i}(d-\tilde{\Delta}_i-s_i)\moi\mphs_i\,.
\end{equation}

We begin with an infinitesimal variation on the boundary quantities
\begin{equation}\label{deltafields}
  (h_{ab},\mphs_i)\rightarrow(h_{ab}+\delta h_{ab},\mphs_i+\delta\mphs_i)\,.
\end{equation}
Then we obtain
\begin{equation}\label{deltas2}
  \delta{S}|_{\text{on-shell}}=\int_{r\rightarrow\infty}\td^dx\left[\frac{\delta {S}}{\delta h_{ab}}\delta h_{ab}+\sum_i\frac{\delta {S}}{\delta\mphs_i}\delta\mphs_i\right]\,,
\end{equation}
which, by definition, gives
\begin{equation}\label{deltas3}
  \delta{S}|_{\text{on-shell}}=-\Omega_{d-1}\left[\tau_{ab}\delta h^{ab}/2+\sum_i\moi\delta \mphs_i\right]\,.
\end{equation}
Here we have used the fact that the boundary is flat and homogenous. Now let us focus on the scaling transformation $(t,x^A)\rightarrow (\lambda^{-1} t, \lambda^{-1} x^A)$ inherited from the scaling symmetry of the AdS spacetime. From the bulk point of view, it means that there are no logarithmic terms in the asymptotical expansion of both metric and matter fields at the AdS boundary. Under the infinitesimal form $\lambda=e^{\varepsilon}$ with $0<\varepsilon\ll1$, we have
\begin{equation}\label{infdeltahphi}
  \delta h^{ab}=-2\varepsilon h^{ab},~~\delta \mphs_i=\varepsilon(d-\tilde{\Delta}_i-s_i)\mphs_i\,.
\end{equation}
Since this scaling transformation is a symmetry of the action, we have $\delta{S}=0$ and obtain Eq.~\eqref{bdtabb2}. A similar result for the scalar field under Euclidian signature was discussed in Refs.~\cite{Petkou:1999fv,Skenderis:2002wp}.

As we have argued that the basic holographic dictionary~\eqref{gkpw} ensures $\moi\mphs_i\geq0$ in a thermodynamically stable phase of a system where there is no spontaneous condensation or there is a spontaneous $\mathbb{Z}_2$-symmetry breaking. From Eq.~\eqref{bdtabb2} one finds that the trace of boundary stress tensor should be non-negative, \emph{i.e.}
\begin{equation}\label{tracebk}
\tau\geqslant 0\,,
\end{equation}
once $d-\tilde{\Delta}_i-s_i>0$.
It comes as a surprise and could be an important feature of a strongly coupled system. On the one hand, the trace of the energy-momentum tensor was proved to be non-positive in a field theory when the interaction is negligible~\cite{PhysRevD.88.125005}. On the other hand, Zel'dovich argued that in fluid matter the strong interaction may raise a positive trace of energy-momentum tensor~\cite{Zeldovich}. Moreover, the positive trace may appear in ultra-strongly coupled systems, for example, the core of neutron stars~\cite{PhysRevD.98.064057,9780387335438}. Under general conditions, we now show that the basic holographic dictionary offers a strong constraint on the trace of the energy-momentum tensor. It not only provides a criterion for judging whether a strong coupling system has a holographic dual description, but also uncovers a potentially important property of some strongly coupled systems. \\

\section{Monotonicity of thermodynamic quantities}
For a CFT, Refs.~\cite{Kovtun:2008kw,Kovtun:2008kx} considered the normalized entropy density $\tilde{c}$ defined as the ratio of entropy density $\mathfrak{s}$ over $T^{d-1}$:
\begin{equation}\label{defcs1}
\tilde{c}=\mathfrak{s}/(T^{d-1}\gamma_d)\,,
\end{equation}
with $\gamma_d$ a constant.
In order to consider a general case  beyond CFT, we introduce the ``normalized grand potential density'' $\g_0$ that is given by
\begin{equation}\label{deffc1}
  \psi=-d^{-1}\gamma_d T^d\g_0(T,\mphs_i)\,.
\end{equation}
We point out that $\g_0$ becomes a constant and reduces to $\tilde{c}$ for a CFT. Moreover, as shown in Ref.~\cite{Kovtun:2008kw}, for a CFT with central charge $c$, $\g_0=c$ is a necessary condition for that  a CFT has a dual gravitational description.

When a CFT is deformed by some relevant operators, both $\g_0$ and $\tilde{c}$ cease to be constant. Using $\mathfrak{s}=-\partial \psi/\partial T$, we can find that
\begin{equation}\label{cts2}
  \psi d=-T\mathfrak{s}+T^{d+1}\gamma_d d^{-1}\frac{\partial\g_0}{\partial T}\,,
\end{equation}
where $\mathfrak{s}$ is the entropy density.

Let us use the scaling hypothesis which stipulates that the grand potential density is a homogeneous function of its thermodynamic quantities. Here we consider the case of the systems having no Weyl anomaly, for which the scaling symmetry of the bulk fields guarantees that the dual boundary theory has the scaling symmetry $(\psi,T,\mphs_i)\rightarrow (\lambda^d\psi,\lambda T, \lambda^{d-\tilde{\Delta}_i} \mphs_i)$  with $\lambda$ a positive constant. Then, Euler's homogeneous function theorem yields
\begin{equation}\label{scaling3}
  \psi d=-T\mathfrak{s}-\sum_{i}(d-\tilde{\Delta}_i) \mphs_i\moi\,.
\end{equation}
For a system that has no spontaneous condensation or has a spontaneous $\mathbb{Z}_2$-symmetry breaking, we have shown that $\moi\mphs_i\geqslant0$~\eqref{os} in a thermodynamically stable phase. Therefore, one immediately obtains from Eq.~\eqref{scaling3} that
\begin{equation}\label{dgdT}
  \partial\g_0/\partial T\leqslant0\,.
\end{equation}
\emph{i.e.} $\g_0$ is a non-increasing function of $T$. Moreover, using Eqs.~\eqref{defcs1}-\eqref{cts2}, we obtain
\begin{equation}\label{eqforcf}
  \tilde{c}=\g_0+d^{-1}T\frac{\partial\g_0}{\partial T}\leqslant \g_0\,.
\end{equation}
Therefore, for a strongly coupled system that has a dual gravitational description, if it has no spontaneous condensation or has a spontaneous $\mathbb{Z}_2$-symmetry breaking, $\g_0$ must be a non-increasing function of $T$ and should satisfy $ \tilde{c}\leq\g_0$. Let us consider the high-temperature limit, $T\rightarrow\infty$, for which other energy scales become irrelevant and therefore the conformal symmetry will restore. We denote the central charge in this limit to be $c_{\text{UV}}$. Then the normalized entropy density equals to the central charge~\cite{Kovtun:2008kw}, \emph{i.e.} $\tilde{c}|_{T\rightarrow\infty}=c_{\text{UV}}$. Note also that in the high-temperature limit $\g_0|_{T\rightarrow\infty}=c_{\text{UV}}$. We then find that
\begin{equation}
\g_0\geq c_{\text{UV}}\,.
\end{equation}
It is still a longstanding issue to quantify the number of effective degrees of freedom of a system as a function of its energy scale. In particular, the $c$-theorem~\cite{Zamolodchikov:1986gt,Komargodski:2011vj} hasn't yet been extended to a general case at all energy scales. Here we provide a candidate $c$-function $\g_0$ which is monotonic as one accesses lower scales and potentially gives a clear measurement for the number of effective degrees of freedom at a given temperature.\\

\section{Discussion}
\subsection{Paradox of negative susceptibility}
We now discuss a ``paradox'' raising by the negative susceptibility, for which the resolution begs a new question on the basic dictionary~\eqref{gkpw} but so far has received limited attention. Although our following discussion will focus on the scalar model, a similar phenomenon will also appear in other fields.

We consider a simple model which describes a bulk free scalar field $\varphi$ in $(d+1)$-dimensions.
\begin{equation}\label{lmlead1}
  \mathcal{L}_m=-\frac{1}{2}\partial^\mu\varphi\partial_\mu\varphi-\frac{1}{2}m^2\varphi^2\,.
\end{equation}
This describes a scalar operator $\mathfrak{o}$ at the boundary with the conformal dimension $\Delta$. We consider the case for which the backreaction of the scalar to the background geometry can be ignored.
In the following discussion, we focus on the case
$$\Delta\in(d/2-1,d/2)\cup(d/2,d)\,,$$
due to the unitarity bound and the requirement that the source will not destroy the asymptotically AdS geometry as $r\rightarrow\infty$. For simplicity, we do not consider the one that saturates the BF bound with $\Delta=d/2$.

When $\Delta\in(d/2-1,d/2)$, we have to consider the alternative quantization by choosing $\mphe$ as the source, \emph{i.e.} $J=\mphe$. From the bulk point of view, it corresponds to  $\nu\in(0,1)$ and $\Delta=d/2-\nu$ with $\nu=\sqrt{d^2/4+m^2}$. Then we have $$\langle\mathfrak{o}\rangle=(2\Delta-d)\mphs=-2\nu\mphs\,.$$
The relationship between $\mphs$ and $\mphe$ can be found by solving the equation $\nabla^2\varphi=m^2\varphi$ under the background of Schwarzschild-AdS black brane.
More precisely, the solution of $\varphi(r)$ can be expressed in terms of the hypergeometric function, from which the susceptibility reads
\begin{equation}\label{negsus}
\frac{\partial \langle\mathfrak{o}\rangle}{\partial\mphe}=(d-2\Delta)\left(\frac{d}{4\pi T}\right)^{2\Delta-d} K(d/2-\Delta,d)\,,
\end{equation}
with
\begin{equation}\label{phiephis2}
\begin{split}
  K(x,d)&=\frac{\Gamma(1/2-x/d)^2\Gamma(1+2x/d)}{\Gamma(1/2+x/d)^2\Gamma(1-2x/d)}\,.
  \end{split}
\end{equation}
Since $\Delta\in(d/2-1,d/2)$, one sees that susceptibility is positive, as required by Eq.~\eqref{sucepz1}.

When $d/2<\Delta<d$, we must take the standard quantization and choose $J=\mphs$. The scaling dimension of the operator $\mathfrak{o}$ is $\Delta=d/2+\nu$. One obtains
$$\langle\mathfrak{o}\rangle=(2\Delta-d)\mphe=2\nu\mphe\,,$$
from which one finds
\begin{equation}\label{negsus2}
\frac{\partial \langle\mathfrak{o}\rangle}{\partial\mphs}=(d-2\Delta)\left(\frac{d}{4\pi T}\right)^{2\Delta-d} K(d/2-\Delta,d)\,.
\end{equation}
One then immediately finds that the susceptibility is negative because now $\Delta\in(d/2,d)$. This shows a paradox since the basic relationship~\eqref{gkpw} requires that the susceptibility should be nonnegative.

We stress that such negative susceptibility of standard quantization does not result from the ``semiclassical approximation'' when using the dictionary~\eqref{gkpw}, since the negative susceptibility is always order $\mathcal{O}(1)$ in those models even in the large-$N$ and weak gravitational coupling limit. Moreover, this paradox cannot be relaxed even if one considers the backreaction, since all such models will reduce into the probe free scalar model when the source is infinitesimal (see, \emph{e.g.}~\cite{Li:2020spf}). For the same reason,  this paradox will also appear in the top-down models for which the mass term of~\eqref{lmlead1} is typically replaced by a suitable potential term from a UV complete theory (see, \emph{e.g.} the supergravity model of Ref.~\cite{TOWNSEND198455}).

\subsection{Resolution of the paradox}
The free scalar field shows up in many string theory compactifications, and the probe limit can be considered when the scalar field appears as an excitation on probe D-branes. Since most string compactifications are believed to allow holographic dictionaries, the resolution of this paradox is necessary and important.

The key point is that we have implicitly assumed that $\moi$ defined via Eq.~\eqref{expact10} should be finite at the thermal equilibrium state. However, this is not always true in quantum field theory.
Let's now make some discussion on this assumption. From the definition of the two-point (connected) correlation function $G(x,y)$, we have the following relationship between the expected value and the external source
\begin{equation}\label{onshello1}
  \langle\mathfrak{o}(x)\rangle=\int G(x,y)\mphs(y)\td^dy\,.
\end{equation}
To ensure that $\langle\mathfrak{o}(x)\rangle$ is finite for arbitrary bounded source, it is necessary and sufficient that
\begin{equation}\label{onshello2}
  \int |G(x,y)|\td^dy<\infty\,.
\end{equation}
\emph{i.e.} the correlation function should be absolutely integrable.

If the condition~\eqref{onshello2} is satisfied, the expectation value will be finite and so its derivative with respect to the source is well-defined. Therefore, the proof from Eq.~\eqref{homoz1} to Eq.~\eqref{sucepz1} makes sense and the susceptibility will be nonnegative. However, if the condition~\eqref{onshello2} is violated, $\langle\mathfrak{o}(x)\rangle$ computed from~\eqref{onshello1} could be divergent. In this case, the susceptibility obtained from holography does not correspond to the value appearing in Eq.~\eqref{sucepz1} since $\langle(\mO_i-\langle\mO_i\rangle)^2$ will also be infinite in general. Instead, we should understand it in the following way.
\begin{equation}\label{sucepz1s2}
\begin{split}
  &\left.\frac{\partial\langle\mathfrak{o}_i\rangle}{\partial\mphs_i}\right|_{\mathrm{holography}}\\
  =&\mathrm{analytical~continuation~or~}\\
  &\mathrm{renormailization~of~}
  \frac{\langle(\mO_i-\langle\mO_i\rangle)^2\rangle}{T\Omega_{d-1}}\,.
\end{split}
\end{equation}
Although $\langle(\mO_i-\langle\mO_i\rangle)^2\rangle$ is formally positive-definite, its analytical continuation or renormalization could be negative. For example, the Riemann-Zeta function $\zeta(s)$ is formally defined as $\zeta(s)=\sum_{n=1}^\infty n^{-s}$, which is positive when it converges. However, its analytical continuation of $s=-1$ reads $\zeta(-1)=-1/12<0$.

We now return to our scalar model. From the viewpoint of holography, when the two boundary points are sufficiently close to each other, their correlation cannot ``feel'' the bulk interior and so the correlation will be dominated by the near boundary geometry. As the result, their correlation will be given by the form in AdS vacuum. When they are separated far away, they will probe the black hole geometry near the event horizon. Therefore, the correlation function will be dominated by thermal fluctuation, which in general will show an exponential decay. Thus, the correlation function of the boundary theory satisfies the following universal property.
\begin{equation}\label{asympg1}
  G(x,y)\propto\left\{
  \begin{split}
  &\frac1{|x-y|^{2\Delta}},~~~T|x-y|\ll 1\,,\\
  &\mathrm{decay~expotentially},~~T|x-y|\gg1\,.
  \end{split}
  \right.
\end{equation}
It is now clear that the correlation function is ``absolutely integrable'' if and only if $\Delta<d/2$. One can then conclude that the susceptibility must be nonnegative for the scalar field case if its scaling dimension is less than $d/2$.

When the scaling dimension $\Delta>d/2$, the holographic results should be understood as the analytical continuation from $\Delta<d/2$ to $\Delta>d/2$. We have already obtained the analytical result for $\Delta<d/2$ (see Eq.~\eqref{negsus}), \emph{i.e.}
\begin{equation}\label{negsus2}
\chi(\Delta)=(d-2\Delta)\left(\frac{d}{4\pi T}\right)^{2\Delta-d} K(d/2-\Delta,d)\,,
\end{equation}
which is an analytical function of $\Delta$ and is well-defined even when $\Delta>d/2$. The uniqueness of analytical continuation implies that, after a suitable analytical continuation to remove the divergency of $\langle\mathfrak{o}\rangle$, the resulting susceptibility for $\Delta>d/2$ must still be given by the expression~\eqref{negsus2}. Thus, we obtain that the susceptibility, in this case, is negative.

To further support the above discussion, we consider the BTZ black hole as an example. The (Euclidean) thermal correlation function reads $G(x,y):=g(\rho,\tau)$ with
\begin{equation}\label{gxybtz}
  g(\rho,\tau):=\left(\frac{\pi}{\beta}\right)^{2\Delta}\frac{c_0}{\left(\sinh^2\frac{\pi \rho}{\beta}+\sin^2\frac{\pi\tau}{\beta}\right)^{\Delta}}\,.
\end{equation}
Here $x=(x_1,\tau_1), y=(x_2,\tau_2)$, $\rho=x_1-x_2$, $\tau=\tau_1-\tau_2$ and $\beta=1/T$. The parameter $c_0$ is a positive factor. One finds that this correlation function satisfies the behavior of Eq.~\eqref{asympg1}. The expectation value of the scalar operator then reads
\begin{equation}\label{expscalbtz1}
  \langle\mathfrak{o}(x_1,\tau_1)\rangle=\int g(x_1-x_2,\tau_1-\tau_2)\mphs(x_2,\tau_2)\td x_2\td\tau_2\,.
\end{equation}
In the homogenous case, $\mphs(x_2,\tau_2)$ is constant, for which we have
\begin{equation}\label{expscalbtz2}
  \langle\mathfrak{o}\rangle=\mphs\int g(r,\tau)\td r\td\tau\,.
\end{equation}
We then obtain the susceptibility that is given by
\begin{equation}\label{expscalbtz3}
  \chi(\Delta)=\int g(\rho,\tau)\td \rho\td\tau=\left\{
  \begin{split}
  &\chi_0,~~~~~~\Delta<1\,,\\
  &+\infty,~~~\Delta\geq 1\,.
  \end{split}
  \right.
\end{equation}
with $\chi_0$ a finite positive number that depends on $\Delta$.
For the case $\Delta<1$, we can choose the normalized factor $c_0$ so that the holographic result coincides with the integration~\eqref{expscalbtz2}. For the case $\Delta\geq1$, although $\langle\mathfrak{o}\rangle$ is formally defined by the integration~\eqref{expscalbtz2}, its numerical value is ill-defined. One could treat $\langle\mathfrak{o}\rangle$ as the analytical function of $\Delta$ and make an analytical continuation from $\Delta<1$ to $\Delta>1$. As a consequence, a formally positively defined susceptibility now becomes to be a negative value.

The resolution of the above ``paradox'' raises another interesting issue. While the expectation value computed from the field theory side could be divergent, the holographic computation yields a finite result. This suggests that in quantum field theory the correct partition function associated with the scalar operator should be
\begin{equation}\label{gkpwnew}
 Z_{\text{QFT}}=\lim_{\varepsilon\rightarrow0}\left\langle\exp\sum_i\int_\varepsilon[\mphs_i\mathfrak{o}_i-C(\mphs_i)]\td^dx\right\rangle_{\text{QFT}}\,,
\end{equation}
so as to match holography. Here $\varepsilon$ is a suitable cut-off that regularizes the divergency when $\Delta>d/2$. $C(\mphs_i)$ is a function of the source $\mphs_i$ and cancels the divergency of $\int\mphs_i\mathfrak{o}_i\td^dx$ when $\Delta>d/2$. $C(\mphs_i)$ should also satisfy $\lim_{\varepsilon\rightarrow0}C(\mphs_i)=0$ when $\Delta<d/2$.  Thus far, the details of this new ``counter-term'' $C(\mphs_i)$ are not clear to us, but it must be a nonlinear function. Based on Eq.~\eqref{linearzf1} of appendix~\ref{magsuscp}, the susceptibility could be nonpositive due to the appearance of nonlinear counter-term $C(\mphs_i)$. It is worth having a deeper understanding of this issue in the future.

\section{Summary}
We have shown some necessary conditions for a strongly coupled system that allows a gravity dual description. More precisely, for the case where the correlation function is absolutely integrable, we have uncovered that the trace energy condition should be violated once the scaling dimension $\tilde{\Delta}$ of the operator  $\mathfrak{o}_i$ and its rank $s_i$ satisfy $d-\tilde{\Delta}_i>s_i$. Moreover, we have found a normalized grand potential density $\g_0$ that is a monotonically decreasing function of $T$ and is larger than the central charge of the UV limit. There is an interesting paradox associated with negative susceptibility, for which we have discussed the origin of such paradox and its resolution.

In the present study, we have limited ourselves to a boundary system that is relativistic. Nevertheless, our discussion can be generalized to some non-relativistic theories.  In particular, in Lifshitz holography~\cite{Chemissany:2014xsa,Taylor:2015glc},  the temporal and spatial directions are scaled in a different way $(t,x^A)\rightarrow (\lambda^zt,\lambda x^A)$ with $z$ the dynamical exponent. Such a system is dual to an asymptotically Lifshitz black brane. A similar discussion can be applied to those non-relativistic theories.
~\\


\begin{acknowledgments}
We thank E.~Kiritsis, J.~Zaanen and S.~A.~Hartnoll for helpful conversations. This work was partially supported by the Natural Science Foundation of China Grants No. 12122513, No. 12075298, No.11991052, No.12047503, No.11821505, No.11851302 and No.12005155, and by the Key Research Program of the Chinese Academy of Sciences (CAS) Grant No. XDPB15, the CAS Project for Young Scientists in Basic Research YSBR-006 and the Key Research Program of Frontier Sciences of CAS.
\end{acknowledgments}
\appendix
\section{Discussion about expectation value}\label{expectval}
In this appendix, we will provide a detailed discussion on the expectation value, in particular, Eq.~\eqref{expact1}. Consider a  quantum field theory in $d$-spacetime dimensions and denote the field operator to be $\varpi$. Suppose that an external source $\mphs$ couples with an operator $\mathfrak{o}$. In general, for a thermal equilibrium system, we can always write down the ``first law''
\begin{equation}\label{firstlaw1}
  \td\psi=-\mathfrak{s}\td T-\sigma\td\mphs-\cdots\,,
\end{equation}
where
\begin{equation}\label{defsigma}
  \sigma:=-\partial\psi/\mphs\,.
\end{equation}
The quantity $\sigma$ may be different from the expectation value $\langle\mathfrak{o}\rangle$, since in path integral formulism the expectation value $\langle\mathfrak{o}\rangle$ is defined according to Eq.~\eqref{expact10}, \emph{i.e.}
\begin{equation}\label{expact10b}
 \langle\mathfrak{o}\rangle:=\left.\int\mathfrak{D}[\varpi] \mathfrak{o} e^{-S[\varpi,\mphs]}\right/\int\mathfrak{D}[\varpi] e^{-S[\varpi,\mphs]}\,.
\end{equation}
If the action $S[\varpi,\mphs]$ has the following form
\begin{equation}\label{actionomphs}
  S[\varpi,\mphs]=S_0[\varpi]-\int\mathfrak{o}\mphs\td^dx\,,
\end{equation}
one can prove that
\begin{equation}\label{defsigma2}
  \sigma=\langle\mathfrak{o}\rangle\,.
\end{equation}
Nevertheless, if $S[\varpi,\mphs]$ has a different form from~\eqref{actionomphs}, the result of~\eqref{defsigma2} will become not valid. For example, consider
\begin{equation}\label{actionomphs2}
  S[\varpi,\mphs]=S_0[\varpi]-\int(\mathfrak{o}\mphs+\lambda_1\mathfrak{o}^2\varphi^{(s)}+\lambda_2\mathfrak{o}^2\varphi^{(s)2})\td^dx\,,
\end{equation}
we have $\sigma\neq\langle\mathfrak{o}\rangle$.

Thus,  to obtain Eq.~\eqref{expact1}, we have implicitly assumed that the operator $\mathfrak{o}$ and its ``external source'' $\mphs$ couple with each other in the following way
\begin{equation}\label{couple}
  S[\varpi,\mphs]=S_0[\varpi]-\int(\mathfrak{o}\mphs)\td^dx\,,
\end{equation}
as shown in Eq.~\eqref{gkpw}.
This assumption is nontrivial when we apply the holographic duality to the strong coupling systems. Note that in many materials the external source can contribute to the partition function in a different way, see appendix~\ref{magsuscp} for more details.

\section{Susceptibility in inhomogenous case}\label{suscep}
In the main text, we have argued that the holographic dictatory~\eqref{gkpw} requires a nonnegative susceptibility for homogenous thermal equilibrium states. In this appendix, we will show that this result can be generalized to an inhomogenous state. For simplicity, we only consider a single operator that couples with its source $\mphs$ by~\eqref{couple}. The expectation value of $\mathfrak{o}(x)$ can be obtained by
\begin{equation}\label{expo1}
 \langle\mathfrak{o}(x)\rangle=\frac{\delta\ln Z_{\text{QFT}}}{\delta\mphs(x)}\,.
\end{equation}
In following we will denote
$$\langle\cdots\rangle:=\frac1{Z_{\text{QFT}}}\left\langle\cdots\exp\int\mphs\mathfrak{o}\td^dx\right\rangle_{\text{QFT}}\,.$$
If we define
$${\tilde{O}}:=\int\mathfrak{o}(x)\td^dx\,,$$
then the ``total expectation value'' is given by
\begin{equation}\label{deftotalO}
  \langle{\tilde{O}}\rangle=\left\langle\int\mathfrak{o}(x)\td^dx\right\rangle=\int\langle\mathfrak{o}(x)\rangle\td^dx\,.
\end{equation}
Here we have used the fact that $\langle a+b\rangle=\langle a\rangle+\langle b\rangle$.

We now consider the ``susceptibility'', which is defined as the following functional derivative:
\begin{equation}\label{defsuscep1}
  \chi(x):= \frac{\delta}{\delta\mphs(x)}\langle{\tilde{O}}\rangle\,.
\end{equation}
It is straightforward to show that
\begin{equation}\label{defsuscep12}
\begin{split}
  \chi(x)&=\int\td^dy\frac{\delta^2}{\delta\mphs(x)\delta\mphs(y)}\ln Z_{\text{QFT}}\\
  &=\int\td^dy\left[\langle\mathfrak{o}(y)\mathfrak{o}(x)\rangle-\langle\mathfrak{o}(x)\rangle\langle\mathfrak{o}(y)\rangle\right]\\
  &=\langle{\tilde{O}}\mathfrak{o}(x)\rangle-\langle{\tilde{O}}\rangle\langle\mathfrak{o}(x)\rangle\,.
  \end{split}
\end{equation}
The function $\chi(x)$ can be either positive or negative somewhere. Nevertheless, its average on the whole Eucliadean spacetime, \emph{i.e.}
\begin{equation}\label{valuebarchi}
\begin{split}
  \bar{\chi}&=\int\chi(x)\frac{\td^dx}{V}=\int[\langle{\tilde{O}}\mathfrak{o}(x)\rangle-\langle{\tilde{O}}\rangle\langle\mathfrak{o}(x)\rangle]\frac{\td^dx}{V}\\
  &=\frac{\langle{\tilde{O}}^2\rangle-\langle{\tilde{O}}\rangle^2}{V}\,
  \end{split}
\end{equation}
must be nonnegative. Here $V:=\int\td^dx$. In the homogenous and thermal equilibrium case, we have $\bar{\chi}=\partial\langle\mathfrak{o}\rangle/\partial\mphs$, $V=T^{-1}\Omega_{d-1}$ and $\tilde{O}=T^{-1}\mO$. Then Eq.~\eqref{valuebarchi} just reduces to Eq.~\eqref{sucepz1}.

\section{Example of negative susceptibility}\label{magsuscp}
After showing the nonnegativity of susceptibility, one may have some confusion. For example, if we treat $\mphs$ as the external magnetic intensity $B$, it is well known that many materials have negative magnetic susceptibility.

To understand this problem, let us consider the famous ``Landau diamagnetism'' as an example. Though this is the standard context in the textbook of statistic mechanics of magnetic materials, the reader of the holographic duality community might not be familiar with it. Therefore, we make a brief introduction here (for more details, see \emph{e.g.} Ref.~\cite{9780123821881}).

The Landau diamagnetism describes diamagnetism in a free electron gas. In the presence of a uniform external magnetic field $B$ directed along the $z$-axis, a charged particle would follow a helical path whose axis is parallel to the $z$-axis and its projection on the $(x,y)$-plane is a circle.  Quantum-mechanically, the energy associated with the circular motion is quantized and reads
\begin{equation}\label{energyp1}
  \varepsilon_n=\mu_BB(2n+1)+p_z^2/(2m), ~~n=0,1,2,\cdots\,,
\end{equation}
where $\mu_B$ is the Bohr magneton. At the high-temperature limit, the system is effectively Boltzmannian. The partition function in the continuous limit reduces to
\begin{equation}\label{partQf1}
  Z=\exp\left[a_0\Omega_{d-1}B\left(\sum_{n=0}^\infty e^{-\frac{\mu_BB(2n+1)}{T}}\right)\int_{-\infty}^{\infty} e^{-\frac{p_z^2}{2mT}}\td p_z\right]\,.
\end{equation}
Here $a_0$ is a positive constant and its expression can be found in Ref.~\cite{9780123821881}. This gives us
\begin{equation}\label{partQf1}
  Z=\exp\left[a_0\Omega_{d-1}\frac{B{\sqrt{2\pi m/T}}}{2\sinh(\mu_BB/T)}\right]\,.
\end{equation}
From the partition function~\eqref{partQf1} one can find that the susceptibility in the limit $\mu_BB\ll T$ is given by
\begin{equation}\label{negsuspt1}
  \chi=-\frac{\bar{n}\mu_B^2}{3T}<0\,,
\end{equation}
with $\bar{n}$ the particle number density.

It is clear that partition function~\eqref{partQf1} \textit{cannot} be written into a magnetic dipole coupling, \emph{ i.e.},
\begin{equation}\label{linearzf1}
  Z\neq\text{Tr}\exp[-(H_0+Bc_1)/T]\,,
\end{equation}
where both operators $H_0$ and $c_1$ are independent of $B$. Thus, the response of applied magnetic field in a ``Landau diamagnetic'' material can not be described by Eq.~\eqref{gkpw}.

In general, when we turn on the source $y$ for a thermal equilibrium system, the Hamiltonian can always be written into $\mathcal{H}_0+\mathcal{H}_1$, where $\mathcal{H}_1$ is the additional contribution due to the external source $y$. Then the partition function reads $Z=\text{Tr}e^{-(\mathcal{H}_0+\mathcal{H}_1)/T}$, and the response is $X:=T\partial_y\ln Z$.  The susceptibility is given by
\begin{equation}\label{dYdX1}
 \frac{ \partial X}{\partial y}=T^{-1}[\langle (\partial_y\mathcal{H}_1)^2\rangle-\langle \partial_y\mathcal{H}_1\rangle^2]-\langle\partial_y^2\mathcal{H}_1\rangle\,.
\end{equation}
It is now manifest that if $\mathcal{H}_1$ is a linear function of $y$, the stable equilibrium phase should have a nonnegative susceptibility. This is what we have found in the main text. Otherwise, the susceptibility could be negative. Indeed, in many materials, the response to an applied magnetic field is complicated and is not simply described by the local magnetic dipoles. In such cases, the susceptibility can be negative.

\section{Susceptibility and thermodynamic stability}\label{suscepstab}
The ``heat capacity'' and ``minus of compressibility'' are also two kinds of susceptibilities corresponding to temperature and pressure, respectively. It is known that thermodynamic stability requires both the heat capacity and the minus of compressibility to be nonnegative. This usually leads to a widespread misconception: the nonnegativity of susceptibility is always necessary for thermodynamic stability. This appendix aims to clarify this misunderstanding. Particularly, we will explain why the susceptibility in some cases has a relationship to stability but in other cases it does not. In the following, we will use the  heat capacity and the magnetic susceptibility as concrete examples.

\begin{figure}
\centering
\includegraphics[width=0.30\textwidth]{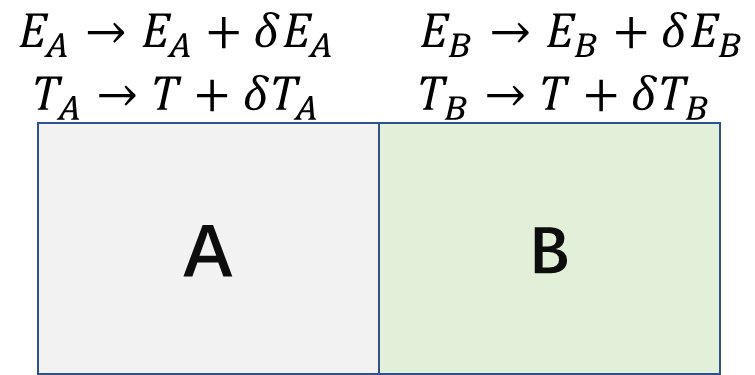}
   \caption{Energy fluctuations in two subsystems}\label{stab1}
\end{figure}

Let us first explain why the negative  heat capacity will lead to instability. This argument can be found in many standard textbooks. We write it here again in order to compare it with the magnetic susceptibility. The specific heat is a  susceptibility of temperature defined as
\begin{equation}\label{defc}
  C=\partial E/\partial T\,.
\end{equation}
There are two characteristic properties that play key roles:
\begin{enumerate}
\item[(1)] The energy is a conserved charge.
\item[(2)] The energy can flow from the high-temperature region into the low-temperature region spontaneously without causing any other change.
\end{enumerate}
Let us consider an isolated system that contains two subsystems A and B as shown in Fig.~\ref{stab1}. Assume that the system is in equilibrium at temperature $T$. Now consider that, due to a fluctuation, the energy of subsystem A becomes $E_A+\delta E_A$ with $\delta E_A>0$ and subsystem B then becomes $E_B+\delta E_B$.
Since the total energy is conserved, we have $\delta E_B=-\delta E_A<0$. Let's consider that the temperature susceptibility $C$ is negative. Therefore, the temperature of subsystem A becomes $T_A=T+\delta T_A<T$ and the temperature of subsystem B becomes $T_B=T+\delta T_B>T$. Since energy will run from the high-temperature region into the low-temperature region spontaneously, more energies will run into A from B. This results in the temperature of subsystem B becoming higher and higher while its energy becomes less and less. Thus, the system is unstable under fluctuation. The same argument will also work for compressibility if one uses ``minus volume ($-V$)'' to replace energy and pressure to replace temperature.

Now let us consider the magnetic susceptibility which is defined as
\begin{equation}\label{defc}
  \chi=\partial M/\partial H\,,
\end{equation}
where $H$ stands for the magnetic field and $M$ stands for the magnetic moment. It is clear that $M$ is not a conserved charge since the magnetic moment can disappear. Moreover, the magnetic moment $M$ does not always flow from the high magnetic field region into the low magnetic region spontaneously. For example, if one puts a magnet into the water and then takes it out after a long time, one will find that the magnetic moment of the magnet will not decrease and the magnetic moment of water will not increase------no magnetic moment runs from magnet into water.  If one follows the above argument of the specific heat, one can find the following differences:
\begin{enumerate}
\item[(1)] The fluctuations of magnetic moment in two subsystems are independent.
\item[(2)] Even if at a special moment with $\delta M_A>0$ and $\delta M_B<0$, the magnetic moment of subsystem B will not run into subsystem A spontaneously without causing any other change.
\end{enumerate}
Therefore, it is easy to see that negative magnetic susceptibility does not cause instability.

These two concrete examples clearly show that the nonnegativity of susceptibility and thermodynamical stability, in general, will not have a close relationship.  There is only one simple situation, where the stability has a relationship to the sign of susceptibility: the source contributes to the action linearly-----this is the situation considered in the holographic formula~\eqref{gkpw}.

We can also understand why the nonnegativity of susceptibility is not required by stability from the 2nd law of thermodynamics. Let's consider the canonical ensemble of which the dynamics is given by free energy $F(T,X)$. The $X$ stands for the extensive independent variable (``minus volume'', particle number, magnetic moment, and so on), and the conjugate intensive quality ( pressure, chemical potential, the magnetic field, and so on) is denoted by $y$. Then one has
\begin{equation}\label{free1stlaw}
  \td F=-S\td T+y\td X\,.
\end{equation}

We begin with a \textit{``wrong'' derivation} on the stability of the equilibrium condition. Since the 2nd law of thermodynamics requires the free energy in an equilibrium state to have minimal value, one has $\delta^2F\geq0$. Considering that the temperature is fixed, one obtains
\begin{equation}\label{delta2f}
  \delta^2F=\frac{\partial^2F}{\partial X^2}(\delta X)^2\geq0\Rightarrow\frac{\partial y}{\partial X}\geq0\,.
\end{equation}
This shows that the susceptibility should be nonnegative. Nevertheless, this is wrong since the equilibrium state also requires $\delta F=0$. Following the logic of \eqref{delta2f}, one should obtain
\begin{equation}\label{delta2f2}
  \delta F=\frac{\partial F}{\partial X}\delta X=0\Rightarrow y=0\,.
\end{equation}
This is obviously wrong. Therefore, Eq.~\eqref{delta2f} is not a correct result.

The correct derivation is as follows. One separates the system into two subsystems A and B. The 2nd law of thermodynamics leads to the following equations on an equilibrium state.
\begin{equation}\label{equilieq1}
  \left\{
  \begin{split}
  &\delta F=\delta F_A+\delta F_B=0,~~~(\text{equilibrium condition})\\
  &\delta^2F=\delta^2 F_A+\delta^2 F_B\geq0,~~~(\text{stable condition})\\
  &C(X_A,X_B)=0,~~~(\text{constraint condition})
  \end{split}
  \right.
\end{equation}
We will show that the constraint condition is also important.

Let us first consider $X=-V$ as the concrete example. Then the variable $y$ stands for the pressure. In flat spacetime, the variation of volume is caused by the move/deformation of boundary between A and B. Thus, the constraint equation reads
\begin{equation}\label{constraintv1}
  C(X_A,X_B)=X_A+X_B-X_0\,.
\end{equation}
Here $X_0$ is a constant and stands for the minus of total volume. This leads to
\begin{equation}\label{deltaxaxb}
  \delta X_A=-\delta X_B,~~\delta^2 X_A=-\delta^2 X_B\,.
\end{equation}
The equilibrium condition then yields
\begin{equation}\label{deltaxaxb2}
  \delta F_A+\delta F_B=y_A\delta X_A+y_B\delta X_B=(y_A-y_B)\delta X_A=0\,.
\end{equation}
This gives us the correct equilibrium condition: the pressures of the two subregions are the same. The stable condition then shows that
\begin{equation}\label{deltaxaxb3}
\begin{split}
  &\delta^2 F_A+\delta^2 F_B\\
  =&y_A\delta^2X_A+y_B\delta^2 X_B+\frac{\partial y_A}{\partial X_A}(\delta X_A)^2+\frac{\partial y_B}{\partial X_B}(\delta X_B)^2\\
  =&\left(\frac{\partial y_A}{\partial X_A}+\frac{\partial y_B}{\partial X_B}\right)(\delta X_B)^2\geq0\,.
  \end{split}
\end{equation}
Here we have used the constraint condition~\eqref{deltaxaxb} and $y_A=y_B$. Now assume that $A$ is the environment and is large enough, \emph{i.e.} $X_A\gg X_B$. Then we have
\begin{equation}\label{envir1}
  \left|\frac{\partial y_A}{\partial X_A}\right|\ll\left|\frac{\partial y_B}{\partial X_B}\right|\,.
\end{equation}
Therefore, the stable condition~\eqref{deltaxaxb3} immediately leads to $\partial y_B/\partial X_B\geq0$. Since $y$ stands for pressure and $X$ stands for $-V$, this gives us the correct stable condition: the minus of compressibility should be nonnegative.

For general variable $X$, such as the magnetic moment, one should \textit{not} expect that the constraint equation $C(X_A,X_B)$ is as same simple as Eq.~\eqref{constraintv1} since $X$ may not be conserved. Then one cannot obtain Eqs.~\eqref{deltaxaxb}-\eqref{deltaxaxb3}, particularly, one \textit{cannot} obtain
\begin{equation}\label{notobtaineq}
  y_A\delta^2X_A+y_B\delta^2 X_B=0\,.
\end{equation}
Therefore, the nonnegative susceptibility is not always guaranteed by the stability of equilibrium.

\section{A brief discussion on alternative quantization}\label{alternative}
In the main text, we only consider the so called ``standard quantization''. The alternative quantization chooses the term $\vec{\varphi}^{(e)}:=(\mphe_1,\cdots,\mphe_{\mathcal{N}})$ to be the source. This corresponds to a Legendre transformation $(T,\vec{\varphi}^{(s)})\rightarrow(T,\vec{\varphi}^{(e)})$, and the corresponding grand potential density becomes
\begin{equation}\label{defF12}
  f(T,\vec{\varphi}^{(e)})={\psi}(T,\vec{\varphi}^{(s)})-\sum_i\zeta_i\mphe_i\,.
\end{equation}
Here $\zeta_i$ satisfies $\mphe_i(\partial\zeta_i/\partial\mphs_i)_T=(\partial{\psi}/\partial\mphs_i)_T$ in order to match the first law
\begin{equation}\label{defF12}
  \td f=-\mathfrak{s}\td T-\sum_i\zeta_i\td\mphe_i\,.
\end{equation}
The expectation values of the operators read $\moi=-\partial f/\partial\mphe_i=\zeta_i$. The scaling dimension of $\langle\mathfrak{o}\rangle$ is $\Delta_i=d-\tilde{\Delta}_i$. Note that we also need to modify the boundary term ${S}_{ct}$ of the bulk action~\eqref{bulkaction} such that the on-shell Euclidian action satisfies $T{S}_{\text{Euclidian,on-shell}}=f\Omega_{d-1}$. Then the holographic dictionary requires
\begin{equation}\label{thersta2}
  \frac{\partial\moi(T,\vec{\varphi}^{(e)})}{\partial\mphe_i}\geq0\,.
\end{equation}
All our discussions in the standard quantization can be applied to the case with the alternative quantization.

\bibliography{inrqbh1-ref}


\end{document}